\documentclass[aps,floats,floatfix,showpacs,amssymb,prd,twocolumn,superscriptaddress,nofootinbib,nolongbibliography,reprint]{revtex4-2}

\usepackage{amssymb,amsmath,verbatim,mathtools,needspace,enumitem,etoolbox,graphicx,physics,microtype,afterpage,bigints,gensymb,tabularx,xspace}

\usepackage[dvipsnames]{xcolor}
\definecolor{linkcolor}{rgb}{0.0,0.3,0.5}
\definecolor{dodgerblue}{HTML}{1E90FF}
\usepackage[unicode, colorlinks=true, linkcolor=linkcolor, citecolor=linkcolor, filecolor=linkcolor,urlcolor=linkcolor, pdfusetitle]{hyperref}
\usepackage[all]{hypcap}
\usepackage[T1]{fontenc}
\usepackage[utf8]{inputenc}
\usepackage{orcidlink}

\usepackage{xcolor}
\definecolor{ochre}{rgb}{0.8, 0.47, 0.13}

\usepackage[normalem]{ulem}

\makeatletter
\newcommand*{\balancecolsandclearpage}{\close@column@grid \cleardoublepage \twocolumngrid}
\makeatother

\begin{document}

\title{The accuracy of merger times in hierarchical black hole triples}

\author{Giulia Fumagalli$\,$\orcidlink{0009-0004-2044-989X}}
\email{gfumagal@caltech.edu}
\affiliation{TAPIR 350-17, California Institute of Technology, 1200 E California Boulevard, Pasadena, CA 91125, USA}

\author{Alejandro Vigna-Gómez$\,$\orcidlink{0000-0003-1817-3586}}
\email{avignagomez@nbi.ku.dk}
\affiliation{Niels Bohr International Academy, Niels Bohr Institute, Blegdamsvej 17, DK-2100 Copenhagen, Denmark}


\begin{abstract}

Orbit-averaged equations are widely used to model the gravitational-wave-driven evolution of compact binaries. Their validity relies on the radiation-reaction timescale being much longer than the orbital period, an assumption that can break down for highly eccentric systems. In this work, we investigate the impact of orbit averaging on merger-time estimates by comparing orbit-averaged and non-orbit-averaged prescriptions. We apply our analysis to two different astrophysically motivated populations of black hole binaries in hierarchical triple systems, where von Zeipel–Lidov–Kozai oscillations can drive the inner binary eccentricities close to unity.
We find that merger-time estimates obtained with orbit-averaged equations remain remarkably accurate across a wide range of eccentricities, even when the orbital timescale exceeds the radiation-reaction timescale. Significant discrepancies arise only for the most extreme and compact binaries, simultaneously characterized by very high eccentricities and small semi-latus recta. In this regime, the merger time becomes strongly dependent on the initial orbital phase, with different phases yielding predictions that can differ by several orders of magnitude. We also compare our results with an orbit-averaged merger-time prescription specifically developed for hierarchical triple systems and find substantial differences between this prescription and both the generic orbit-averaged and non-orbit-averaged calculations in the high-eccentricity regime.
While these results demonstrate the robustness of orbit-averaged merger-time estimates for most systems, they also highlight the potential importance of non-orbit-averaged effects when merger times are used to infer compact-binary delay-time distributions and formation channels.
\end{abstract}

\maketitle

\section{Introduction} \label{intro}
The more than 300 gravitational-wave (GW) detections of stellar-mass binary black hole (BH) mergers reported by the LIGO/Virgo/KAGRA Collaboration \cite{2026arXiv260527225T} are bringing us closer to addressing the still open question of the astrophysical origin of BH binaries. Over the past decade, the number and diversity of proposed formation channels have grown substantially \cite{2021hgwa.bookE..16M,2022PhR...955....1M}. These scenarios are often broadly divided into two families: isolated binary evolution, in which stellar binaries evolve into BH binaries through stellar evolution, and dynamical formation, in which BH binaries are assembled through gravitational interactions in dense stellar environments. Each of these broad classes encompasses a variety of distinct astrophysical subchannels, and some can produce binaries with observational properties compatible with both formation pathways, making their origins difficult to distinguish. Despite the diversity of the proposed channels and the physical processes underlying them, all viable scenarios must satisfy a common requirement: the resulting BH binary must merge within the age of the Universe. 
Binaries that fail to meet this requirement cannot %
contribute to the observed GW population.  %
Accurate estimates of merger times are therefore essential for assessing the viability of formation scenarios and predicting their contribution to the observed merger rate.
The merger time depends sensitively on the binary's orbital separation and eccentricity: highly eccentric and/or tight binaries can merge much more rapidly than wide and/or quasi-circular systems, leading to differences of several orders of magnitude in the predicted merger times. The distribution of delay times between binary formation and merger therefore directly influences the merger rate expected from each formation channel.

Merger times can be also particularly sensitive to the assumptions made about the environment governing binary evolution and, crucially, to the approximations adopted in the treatment of the dynamics \cite{1964PhRv..136.1224P,2018ApJ...863...68L,2007MNRAS.379..956D,2006ApJ...639..999G}.
The standard estimate for the merger time of an isolated binary evolving under GW emission was derived in Ref.~\cite{1964PhRv..136.1224P}. To obtain this estimate, the equations of motion are averaged over an orbital period, filtering out short-timescale dynamics and retaining only the secular evolution of the orbital elements.
While this approach is  %
widely used and provides an accurate description of long-term binary evolution within its regime of validity, it breaks down for sufficiently eccentric and tight binaries \cite{2014ApJ...781...45A,2025PhRvD.112b4012F}. Orbit averaging assumes that the orbital parameters evolve adiabatically, with negligible changes over a single orbital period, and removes the explicit dependence on the orbital phase from the equations of motion. At high eccentricity and relatively small separations, however, the binary can undergo rapid, non-adiabatic evolution near pericenter, violating this assumption. A non-orbit-averaged treatment retains the full phase dependence of the dynamics and can therefore capture these short-timescale effects.

The astrophysical consequences of this breakdown, particularly for merger-time estimates, have not yet been quantified. Here, we quantify this effect by focusing on BH binaries in hierarchical triple systems, where von Zeipel-Lidov-Kozai (ZLK) oscillations can drive the eccentricity to values close to unity \cite{2015MNRAS.447..747L,2018ApJ...863...68L,2022ApJ...934...44M,2023MNRAS.522..937T}, naturally bringing the binary into a regime where the orbit-averaged approximation is known to fail \cite{2025PhRvD.112b4012F}. We consider the populations of BH binaries obtained in Ref.~\cite[VG+25]{2025A&A...699A.272V} and Ref.~\cite[L\&L18]{2018ApJ...863...68L}.
Within these populations, we identify systems for which the inner binary decouples from the outer companion shortly after reaching its maximum eccentricity through ZLK oscillations. We then compute their merger times using the non-orbit-averaged framework introduced in Ref.~\cite{2025PhRvD.112b4012F}, the orbit-averaged prescription of Ref.~\cite{1964PhRv..136.1224P}, the hierarchical-triple-specific merger-time definition of Ref.~\cite{2018ApJ...863...68L}. We find that the orbit-averaged and non-orbit-averaged approaches give comparable merger times for most of the systems analyzed, but can differ by several orders of magnitude for the most extreme binaries, characterized by very high eccentricities and small separations. Finally, we find that the merger-time definition of Ref.~\cite{2018ApJ...863...68L} systematically differs from both the standard orbit-averaged and non-orbit-averaged estimates for the majority of the systems considered.

Although we use hierarchical triples as a concrete astrophysical setting in which extreme eccentricities naturally arise, the orbital-phase effects associated with non-orbit-averaged dynamics are not specific to this formation channel and apply more generally to BH binaries evolving in the highly eccentric regime. Our results therefore provide a broader assessment of when orbital-phase information must be retained to obtain reliable merger-time predictions.

Our paper is organized as follows. In Sec.~\ref{Hierarchical Triple Dynamics and Timescales}, we describe the different timescales on which the dynamics happen and introduce the definitions used for the merger time. In Sec.~\ref{intial conditions}, we describe the BH binary systems considered and the initial conditions chosen for measuring the merger time. In Sec.~\ref{result}, we present our results for the merger time estimates obtained using both orbit-averaged and non–orbit-averaged prescriptions. Finally, in Sec.~\ref{discussion and conclusions}, we discuss the implications of our findings.

\section{Hierarchical Triples: Dynamics and Timescales}
\label{Hierarchical Triple Dynamics and Timescales}

Let us consider a binary composed of two non-spinning BHs with masses $m_1$ and $m_2$, total mass $M = m_1 + m_2$, semi-major axis $a_{\rm in}$, and eccentricity $e_{\rm in}$. Such binary is orbited by a third body of mass $m_3$, forming a hierarchical triple system. In the hierarchical limit, the outer orbit can be treated as a binary between $m_3$ and a point mass $M$ located at the inner binary’s center of mass. We assume the outer orbit to be circular ($e_{\rm out}=0$) and to evolve on timescales much longer than those of the inner binary, allowing us to treat $a_{\rm out}$ as constant (see Ref.~\cite{2026MNRAS.549ag944G} for a treatment of the case in which this assumption is relaxed).

Within this setting, the evolution of the inner binary is governed by the gravitational perturbation from the tertiary companion and by GW emission. The former drives ZLK oscillations through the exchange of angular momentum between the inner and outer orbits, resulting in periodic variations of the magnitude and orientation of the inner orbital angular momentum and, consequently, of the inner-binary eccentricity and inclination \cite{2016ARA&A..54..441N, 2023MNRAS.522..937T}. In the absence of GW emission, the total angular momentum of the triple is conserved, while angular momentum is exchanged between the inner and outer orbits. GW emission, in contrast, dissipates energy and angular momentum from the system, causing the inner binary to shrink and circularize over time \cite{1964PhRv..136.1224P}.

We first introduce the characteristic dynamical timescales of these systems in Sec.~\ref{sec:timescales}. For the populations of BH triples considered here, we use these timescales to identify binaries that violate the assumptions underlying orbit averaging and for which the inner binary can be considered effectively decoupled from the outer companion. We then compute the merger times of the selected systems using the formalisms introduced in Sec.~\ref{Merger Time}.

\subsection{Timescales}\label{sec:timescales}

The evolution of the inner binary is governed by the competition between three characteristic timescales: the orbital period, the timescale of the ZLK oscillations induced by the tertiary, and the radiation-reaction timescale associated with GW emission. The orbital period is
\begin{align}\label{tauorb}
\tau_{\rm orb}
=2\pi\sqrt{\frac{a_{\rm in}^3}{G M}}.
\end{align}
For a circular outer orbit, the characteristic ZLK timescale is \cite{2015MNRAS.447..747L}
\begin{align}\label{tauzlk}
\tau_{\rm ZLK}
=\frac{\tau_{\rm orb}}{2\pi}
\frac{M}{m_3}
\left(\frac{a_{\rm out}}{a_{\rm in}}\right)^3.
\end{align}
This expression is obtained within the orbit-averaged approximation and therefore assumes a separation between the orbital and ZLK timescales, $\tau_{\rm orb}<\tau_{\rm ZLK}$ \cite{2024MNRAS.533..486G,2015MNRAS.452.3610A}.

The third relevant timescale is that associated with GW-driven orbital evolution. We define it in terms of the semi-latus rectum\footnote{We define the radiation-reaction timescale via the evolution of the semi-latus rectum rather than the semi-major axis, as it provides a more robust measure of GW-driven evolution for highly eccentric binaries. In the limit $e_{\rm in}\to1$, $a_{\rm in}\to\infty$ while $p_{\rm in}\to\mathrm{const}$, such that $a_{\rm in}/|\dd a_{\rm in}/\dd t|$ is no longer a well-defined measure of the local radiation-reaction timescale, whereas $p_{\rm in}/|\dd p_{\rm in}/\dd t|$ remains well behaved.},
\begin{align}\label{taurr}
\tau_{\rm rr}
=\frac{p_{\rm in}}
{\left|\dd p_{\rm in}/\dd t\right|}.
\end{align}

We evaluate $\tau_{\rm rr}$ using both orbit-averaged and non-orbit-averaged expressions for $\dd p_{\rm in}/\dd t$. For the former, we use the expression given in Eq.~(48) of Ref.~\cite{2025PhRvD.112b4012F}; for the latter, we use Eq.~(2) of the same reference. Among the available non-orbit-averaged descriptions of eccentric binary dynamics (see, e.g., Ref.~\cite{1990PhRvD..42.1123L}), we adopt the radiation reaction gauge-free, non-adiabatic (NA) formulation of Ref.~\cite{2025PhRvD.112b4012F}. This formulation avoids artificial features associated with gauge-dependent parameters and provides a consistent description of the binary when the orbital evolution is no longer adiabatic. 
In this work we also consider the orbit averaged definition of merger timescale given in Eq. (19) of Ref.~\cite{2020MNRAS.495.2321Z} which account for lower-order PN conservative corrections.
The relative ordering of these timescales determines the dynamical regime of the system. At sufficiently large separations and moderate eccentricities, the system generally satisfies
\begin{align}
\tau_{\rm orb}\ll\tau_{\rm ZLK}\ll\tau_{\rm rr}.
\end{align}
The binary then completes many orbits during a single ZLK cycle, while GW emission modifies the orbital parameters only on a much longer timescale. The dynamics are consequently adiabatic on both the orbital and ZLK timescales, and the evolution can be described within the orbit-averaged approximation \cite{2025PhRvD.112b4012F}.

The situation changes when ZLK oscillations drive the inner binary to high eccentricities. For suitable initial configurations (see Sec.~\ref{intial conditions}), the eccentricity can approach unity \cite{2015MNRAS.447..747L,2018ApJ...863...68L,2022ApJ...934...44M, 2023MNRAS.522..937T}. Because GW emission is strongly enhanced during pericenter passage, the radiation-reaction timescale then decreases rapidly. The system can eventually reach the regime where
\begin{align}
\tau_{\rm rr}\ll\tau_{\rm orb}\ll\tau_{\rm ZLK}.
\end{align}
In this case,  GW-driven evolution occurs on a timescale shorter than the orbital period and the ZLK timescale. The inner binary consequently decouples from the tertiary and proceeds toward merger predominantly under GW radiation reaction, effectively evolving as an isolated binary.

The latter regime lies beyond the domain of validity of the orbit-averaged  equations. We therefore use the non-orbit-averaged formulation of $\tau_{\rm rr}$ to identify binaries entering this regime.

\subsection{Merger Time}\label{Merger Time}
We define the GW-driven merger time of binary BHs as
\begin{align}\label{tm}
t_{\rm m} = \int_{p_{\rm f}}^{p_0} \frac{\mathrm{d} t}{\mathrm{d} p}\,\mathrm{d}p,
 \end{align}
corresponding to the time required for a binary to evolve from an initial semi-latus rectum $p_0$ to $p_{\rm f}=6\,GM/c^2$. Below this separation, the PN approximation adopted in this work ceases to be reliable, and a full numerical-relativity treatment would be required \cite{2009PhRvD..80h4043B}. The additional time required for the binary to evolve from  $6\,GM/c^2$ to merger is expected to provide only a small correction to the total inspiral time and is not included in our definition of $t_{\rm m}$.  %

As with the radiation-reaction timescale defined in Eq.~\eqref{taurr}, the merger time depends on the prescription adopted for $\dd p/\dd t$. Throughout this work, we compute $t_{\rm m}$ using both the orbit-averaged and non-orbit-averaged definitions of  $\dd p/\dd t$ described in Sec.~\ref{timescales}.

For comparison, we also consider the merger-time prescription proposed by Ref.~\cite{2018ApJ...863...68L}, which accounts for the enhancement of GW-driven inspiral  time due to the large eccentricities induced by ZLK oscillations 
\begin{align}\label{tLL} t_{\rm m}^{\rm LL}=t_{\rm m}^{0}(1-e_{\rm max}^{2})^{3}, \end{align} 
where $t_{\rm m}^{0}$ is the merger time of a circular binary, computed from the orbit-averaged equations of Ref.~\cite{1964PhRv..136.1224P}, and $e_{\rm max}$ is the maximum eccentricity attained during the ZLK evolution.

In this work, we adopt a slightly modified definition of $t_{\rm m}^{0}$ by changing the final integration limit. In Ref.~\cite{2018ApJ...863...68L}, $t_{\rm m}^{0}$ is computed by integrating the binaries down to $p_{\rm f}=0$. For consistency with the other merger-time estimates considered here, we instead evaluate $t_{\rm m}^{0}$ using $p_{\rm f}=6\,GM/c^2$. As discussed above, the contribution from the final strong-field regime is expected to provide only a small correction to the merger time.

\subsection{Initial conditions}\label{intial conditions}

\begin{figure}[ht] \includegraphics[width=\columnwidth]{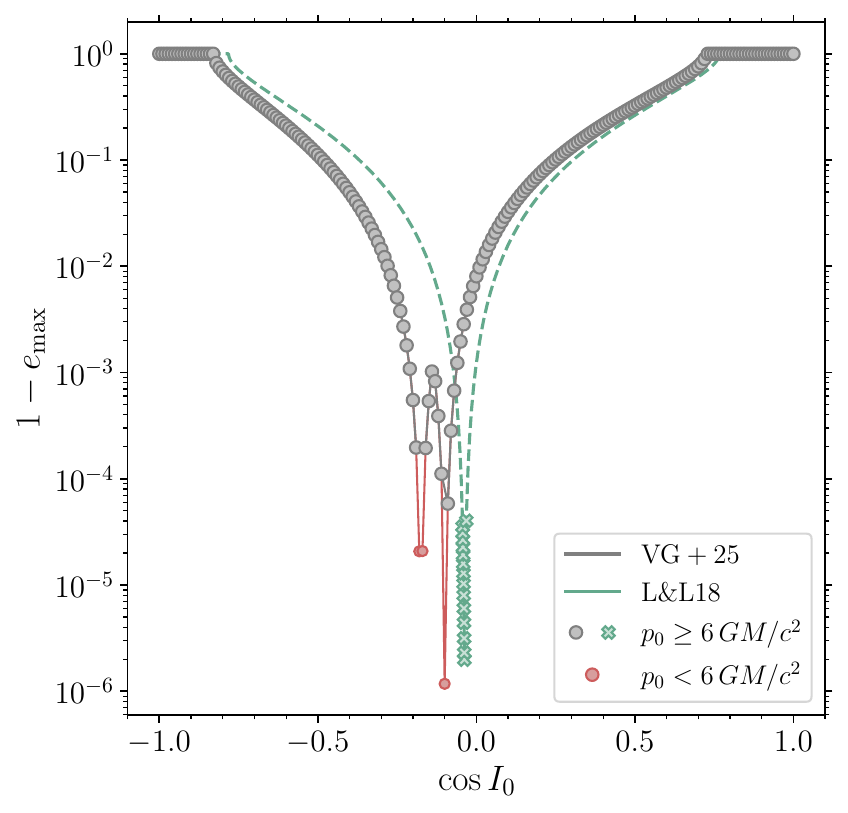}
\caption{Maximum eccentricity of the inner BH binary induced by ZLK oscillations as a function of the initial mutual inclination between the inner and outer orbits. Grey circles show binaries from the population of Ref.~\cite[VG+25]{2025A&A...699A.272V} satisfying $p_0>p_{\rm f}$, while red circles correspond to systems with $p_0\leq p_{\rm f}$. The latter are excluded from our analysis, as their evolution would require numerical-relativity modeling and may lead to direct mergers. Green crosses denote the subset of binaries selected from the population of Ref.~\cite[L\&L18]{2018ApJ...863...68L}.}
\label{cosI0vse} 
\end{figure}

We consider two complementary astrophysical populations of BH binaries in triple systems that undergo ZLK oscillations and can reach high eccentricities \cite{2015MNRAS.447..747L,2022ApJ...934...44M}. The two populations probe different triple configurations: compact, mildly hierarchical systems formed through chemically homogeneous evolution (CHE), and wider, hierarchical systems in which the BH binary is formed first and subsequently perturbed by a tertiary companion.

The first population is taken from Ref.~\cite{2025A&A...699A.272V}. The BH progenitors consist of three non-rotating zero-age main-sequence stars with initial masses $m_1=m_2=m_3=55\,M_\odot$ and metallicity $Z=4.2\times10^{-4}$. The inner binaries are initialized with an orbital period of $1.1$ days and evolved with MESA \cite{}, while the effect of the tertiary is included in post-processing using the analytical treatment of Ref.~\cite{2022ApJ...934...44M}. Provided that no stellar merger occurs, the systems form BH binaries with $m_1=m_2=43.8 M_\odot$, $a_{\rm in}=0.1$ au, $a_{\rm out}=0.5$ au, and an inner orbital period of $1.7$ days.

The second population is drawn from the hierarchical BH triples of Ref.~\cite{2018ApJ...863...68L}, with $a_{\rm in}=100$ au, $a_{\rm out}=4500$ au, and $m_1=m_3=30 M_\odot$, $m_2=20 M_\odot$. In contrast to the CHE population, these systems are substantially wider and the tertiary perturbation is applied to an already formed BH binary. The two populations therefore provide complementary realizations of ZLK-driven eccentric evolution over very different orbital scales.

Figure~\ref{cosI0vse} shows the relation between $e_{\rm max}$ and the initial inclination $I_0$ for the two populations considered in this work. From the population of Ref.~\cite{2018ApJ...863...68L}, we select only a subset of systems, namely those reaching eccentricities larger than those found in the population of Ref.~\cite{2025A&A...699A.272V}. %
We evaluate the timescales defined in Eqs.~\eqref{tauorb}--\eqref{taurr} at the point of maximum eccentricity, characterized by the pair $(a_{\rm in},e_{\rm max})$, or equivalently by $(p_{\rm in}^{},e_{\rm max})$. %
These quantities also define the initial conditions %
we use to evolve the binaries and compute their merger times.
Note that not all binaries in the population of Ref.~\cite{2025A&A...699A.272V} satisfy the requirement $p_{\rm in}>p_{\rm f}$. We discard systems violating this condition, as their evolution would require a numerical-relativity treatment and may correspond to direct mergers.
Finally, when employing the NA equations, an additional initial condition must be specified, namely the true anomaly $f$. This angle takes values in the range $[0,2\pi]$ and describes the instantaneous orbital phase, measured counterclockwise from pericenter.

\subsection{Conversion for NA equations}\label{conversions}

The orbital elements appearing in the NA equations of Ref.~\cite{2025PhRvD.112b4012F} do not directly coincide with the astrophysical quantities $e_{\rm in}$ and $p_{\rm in}$ introduced in the previous section.\footnote{Throughout this work, ``astrophysical'' quantities refer to the orbital parameters provided by population-synthesis models, as described in Refs.~\cite{2025A&A...699A.272V,2018ApJ...863...68L}.} However, the two sets of variables are related through a perturbative transformation.

Specifically, the astrophysical orbital elements $y$ (e.g., eccentricity and semi-latus rectum) are related to the corresponding NA quantities $\bar y$ according to
\begin{align}\label{ytransform}
y = \bar y + \frac{1}{c^5}\,\delta\bar y (\bar y,\alpha,\beta) ,
\end{align}
where $\delta\bar y$ denotes the 2.5PN radiation-reaction correction and depends on the NA variables as well as on the radiation-reaction gauge parameters $\alpha$ and $\beta$.

To initialize the NA evolution using the astrophysical initial conditions defined in Sec.~\ref{intial conditions}, we convert $(p_{\rm in},e_{\rm max})$ into the corresponding barred quantities $(\bar p_{\rm in},\bar e_{\rm max})$. Although the transformation in Eq.~\eqref{ytransform} is analytical, its inverse cannot be expressed in closed form and is therefore computed numerically throughout this work. Explicit expressions for $\delta\bar p$ and $\delta\bar e$ are given in Eqs.~(58)--(59) of Ref.~\cite{2025PhRvD.112b4012F}.

The time coordinate entering the NA equations must be transformed as well. Consequently, all timescales and merger times obtained within the NA framework are mapped back to the astrophysical time coordinate by numerically inverting Eq.~(69) of Ref.~\cite{2025PhRvD.112b4012F}.

The transformation in Eq.~\eqref{ytransform} depends on the choice of radiation-reaction gauge. Throughout this work, we adopt the harmonic gauge, corresponding to setting $\alpha=-1$ and $\beta=0$. This choice follows Ref.~\cite{2025PhRvD.112b4012F} and provides good agreement with the gauge-free radiation-reaction results presented therein.

Finally, following Ref.~\cite{2025PhRvD.112b4012F}, we set $f=\bar f$. Therefore, the astrophysical true anomaly can be used directly within the NA framework without requiring any additional transformation.

\section{Results}\label{result}
\subsection{Timescales comparisons}\label{Timescales comparisons}

 \begin{figure*}[ht]
	\includegraphics[width=\textwidth]{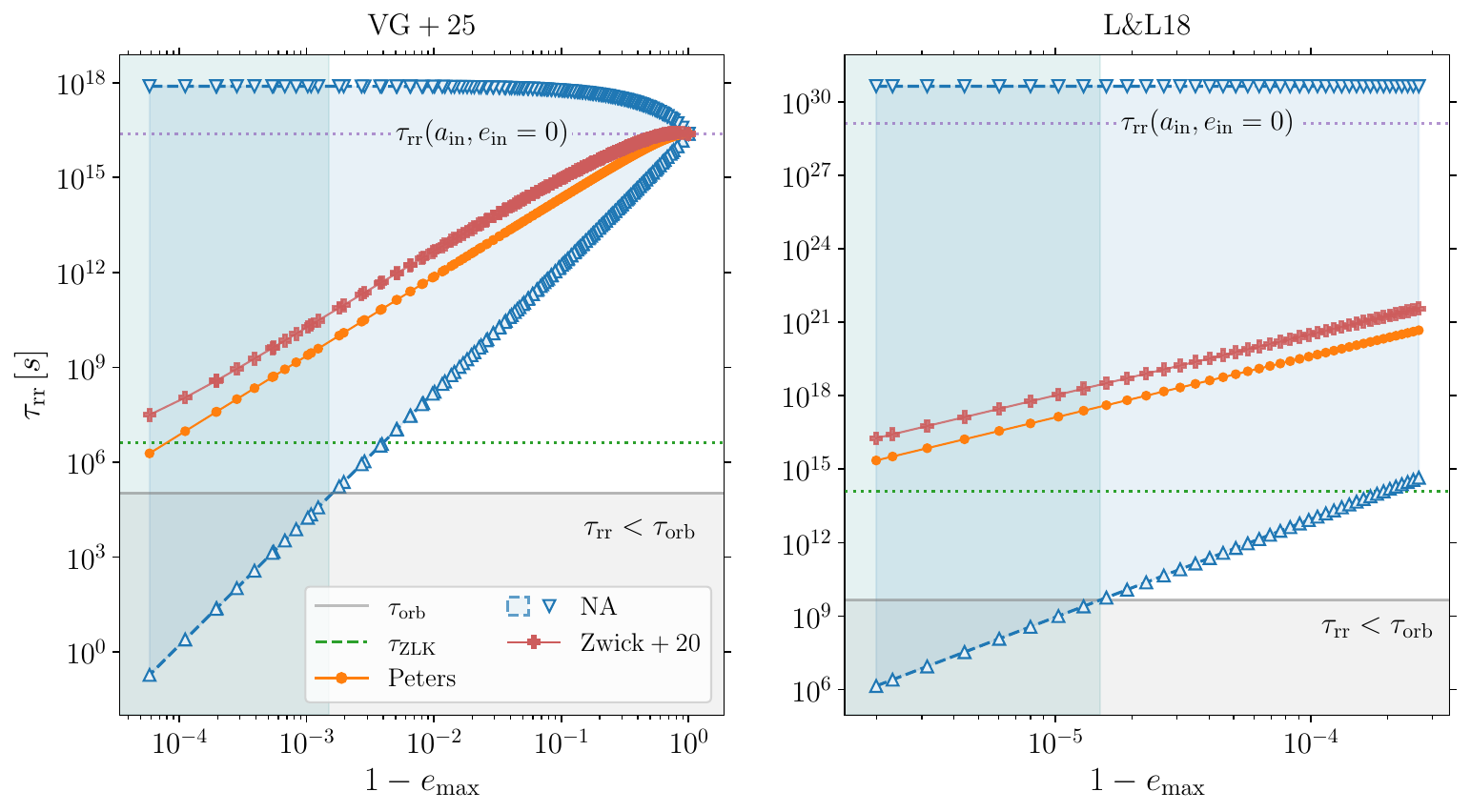}
	\caption{Relevant timescales for the inner BH binaries drawn from the populations of VG+25 (left panel) and L\&L18 (right panel). The radiation-reaction timescale $\tau_{\rm rr}$ is computed using three prescriptions: the standard orbit-averaged formalism based on the equations of Ref.~\cite[Peters]{1964PhRv..136.1224P} (orange circles), the correction to the orbit-averaged prescription proposed by Ref.~\cite[Zwick+20]{2020MNRAS.495.2321Z} (red crosses), and the NA equations. For the latter, upward (downward) light-blue triangles denote the minimum (maximum) value of $\tau_{\rm rr}$, corresponding to $f=0$ ($f=\pi$), while the blue shaded region spans the full range obtained by varying the true anomaly over $f\in[0,2\pi]$. We compare $\tau_{\rm rr}$ with the ZLK timescale $\tau_{\rm ZLK}$ (green dashed line) and the orbital timescale $\tau_{\rm orb}$ (gray line). The gray shaded region marks the regime where $\tau_{\rm rr}<\tau_{\rm orb}$, indicating a breakdown of the timescale hierarchy underlying the secular treatment. Merger times are computed only for binaries whose maximum eccentricity $e_{\rm max}$ lies within the teal shaded region. %
    }
    \label{timescales}
\end{figure*}

Figure~\ref{timescales} shows the characteristic timescales governing the evolution of the systems described in Sec.~\ref{intial conditions} as a function of $(1-e_{\max})$. In particular, we compare the radiation-reaction timescale $\tau_{\rm rr}$ computed using both orbit-averaged and NA equations with the orbital timescale.

For the NA treatment, the initial true anomaly $f$ enters both the conversion from astrophysical to NA orbital elements (Sec.~\ref{conversions}) and the evaluation of the instantaneous radiation-reaction evolution (Eq.~(2) of Ref.~\cite{2025PhRvD.112b4012F}). We therefore show the full range of $\tau_{\rm rr}$ obtained by varying $f$. As expected for an instantaneous quantity, $\tau_{\rm rr}$ depends strongly on the choice of true anomaly, spanning several orders of magnitude between its maximum and minimum values, which occur at $f=0$ and $f=\pi$, respectively. These correspond to the pericenter and apocenter, where the binary reaches its closest and farthest separations and, consequently, its maximum and minimum instantaneous GW emission.

For eccentricities larger than $(1-e_{\max})\simeq1.5\times10^{-3}$, for binaries drawn from the population of VG+25, or larger than $(1-e_{\max})\simeq1.5\times10^{-5}$ for binaries drawn from the population of L\&L18, the NA radiation-reaction timescale can become shorter than both the ZLK timescale $\tau_{\rm ZLK}$ and the orbital timescale $\tau_{\rm orb}$. These thresholds define the upper boundaries of the teal shaded region in Fig.~\ref{timescales}.

 The condition $\tau_{\rm rr}<\tau_{\rm ZLK}$ indicates that GW emission drives the evolution faster than the tertiary companion can modify the inner binary through ZLK oscillations. In this regime, the binary effectively decouples from the outer companion and can be treated as an isolated system. Furthermore, when $\tau_{\rm rr}<\tau_{\rm orb}$, the assumptions underlying orbit averaging break down, eventually making a non-averaged description necessary.

By contrast, the orbit-averaged radiation-reaction timescale remains systematically longer than $\tau_{\rm orb}$ and, in general, also longer than $\tau_{\rm ZLK}$. This behavior is expected, as the orbit-averaged formalism is constructed under the assumption that radiation reaction acts on timescales much longer than the orbital period.

A similar behavior is found for the orbit-averaged radiation-reaction timescale adopted from Ref.~\cite{2020MNRAS.495.2321Z}. In this case, the corrections introduced in that prescription yield larger estimates of $\tau_{\rm rr}$ respect to those obtian with the standard orbit-averaged approach remaining greater than both $\tau_{\rm orb}$ and $\tau_{\rm ZLK}$.

Finally, for binaries with eccentricities lying outside the teal shaded regions of Fig.~\ref{timescales}, the hierarchy $\tau_{\rm orb}<\tau_{\rm ZLK}<\tau_{\rm rr}$ holds, indicating that the orbit-averaged equations remain valid and that the eccentricity evolution continues to be driven primarily by ZLK oscillations.

\subsection{Merger time comparisons}\label{Merger time comparison}
\begin{figure*}[ht]
	\includegraphics[width=\textwidth]{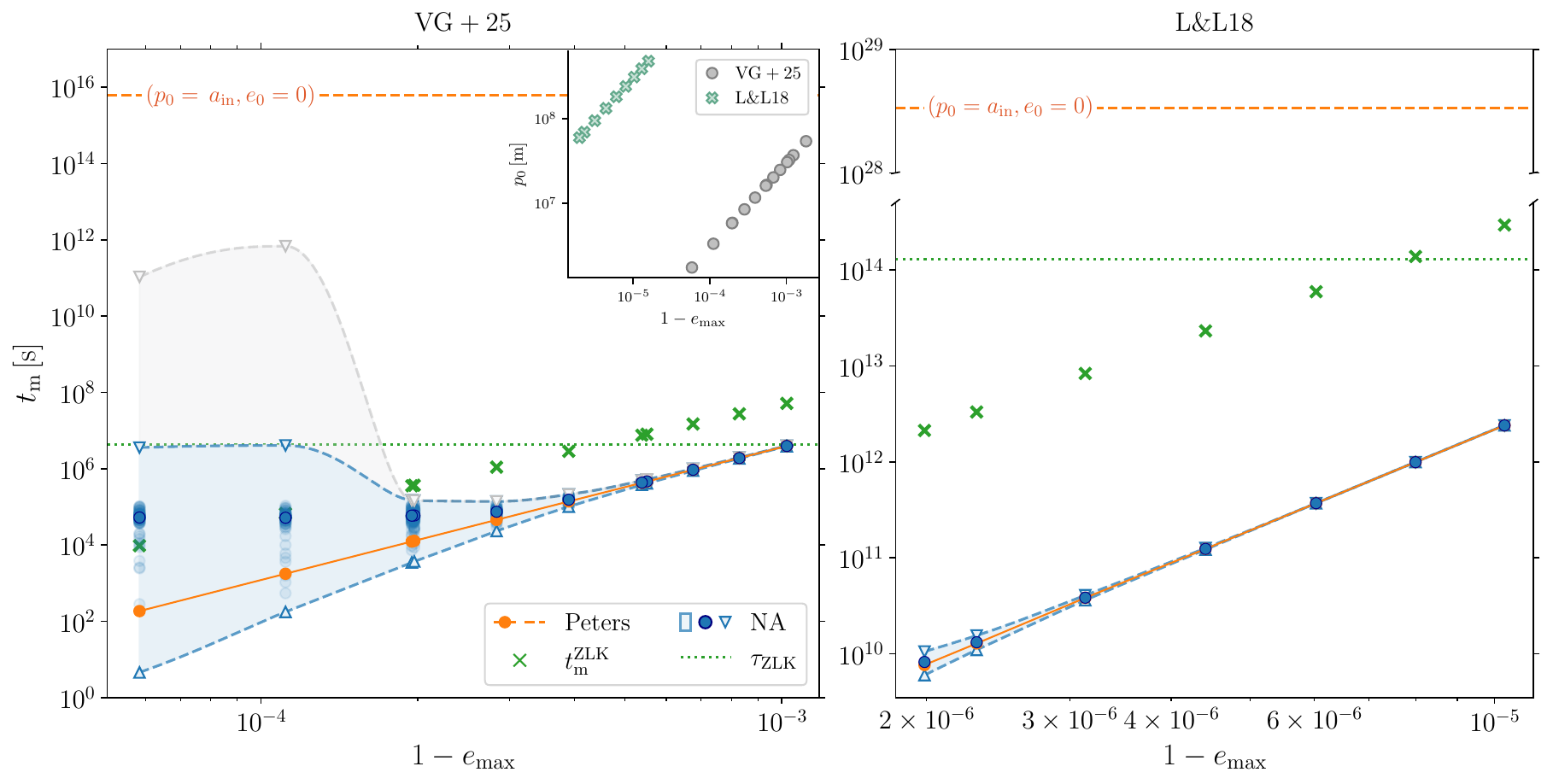} 
	\caption{Time of merger of binaries of VG+25 (left panel) and L\&L18 (right panel) satisfying the timescale hierarchy $\tau_{\rm rr} < \tau_{\rm orb}$, measured using the orbit-averaged prescription of Ref.~\cite{1964PhRv..136.1224P} (orange circles) and Ref.~\cite{2018ApJ...863...68L} (green crosses), and the NA prescription. For the latter, $t_{\rm m}$ is evaluated over 100 initial true anomalies (small light blue circles), sampled from the orbital phase distribution; large blue circles show the corresponding mean. Downward (upward) light blue triangles denote the maximum (minimum) obtained by varying $f$, while the shaded region indicates the full range of possible values.
   Light grey triangles and shaded regions show the values of $t_{\rm m}$ obtained when the evolution of the binary is computed while neglecting the presence of ZLK oscillations.
    The inset in the left panel shows the initial semi-latus rectum of binaries drawn from the two populations considered in this work (gray dots for VG+25 and green crosses for L\&L18). For comparison, the dashed orange line shows the corresponding merger time assuming zero initial eccentricity and employign orbit averged equations.
    }
	\label{merger_time}
\end{figure*}

For binaries satisfying the timescale hierarchy $\tau_{\rm rr} < \tau_{\rm orb} < \tau_{\rm ZLK}$, we compute the merger time as defined in Sec.~\ref{Merger Time}. We consider both the orbit-averaged and  NA definitions of the merger time, as well as the prescription of Ref.~\cite{2018ApJ...863...68L}, and report the corresponding estimates in Fig.~\ref{merger_time}.%
When using the NA framework, we further investigate how different choices of the initial true anomaly affect the merger time. In particular, we consider the values of $f$ that yield the longest and shortest merger times, as well as 100 additional values drawn according to the probability distribution for locating a body along an elliptical orbit with parameters $(e_{\rm in}, p_{\rm in})$, following Ref.~\cite{2025PhRvD.112b4012F}. We also report the arithmetic mean of the resulting $t_{\rm m}$ values, which can be directly compared with the orbit-averaged predictions. 
For binaries with $(1 - e_{\rm max}) > 10^{-3.4}$ belonging to the VG+25 population, the merger-time estimates obtained employing orbit-averaged and NA equations are consistent with each other, and the spread due to different choices of the true anomalies remains limited. For smaller values of $(1 - e_{\rm max})$, significant discrepancies between the two approaches become evident, with orbit-averaged equations underestimating the merger time by up to 3 orders of magnitude with respect to the mean NA evolution.
In this latter regime, the choice of $f$ has a substantial impact, particularly for the two most eccentric binaries. 
In these cases, the inferred merger time can span up to $\sim 6$ orders of magnitude, depending on the initial orbital phase.

Conversely, although binaries drawn from the L\&L18 population have initial eccentricities up to two orders of magnitude larger than those of the VG+25 population, the merger times $t_{\rm m}$ obtained from the orbit-averaged and NA equations are in good agreement. This can be attributed to the larger initial semi-latus recta of the L\&L18 binaries compared with those of the VG+25 population (see Fig.~\ref{merger_time}). The L\&L18 binaries have larger initial semi-major axes (see Sec.~\ref{intial conditions}), which partly compensate for the effect of their higher eccentricities. As a result, the radiation-reaction effects remain sufficiently slow compared with the orbital timescale, mitigating potential non-adiabatic effects.

In both populations, the values of $f$ that yield extrema in the merger time computed with the NA equations differ from those corresponding to the extrema in Fig.~\ref{timescales}. This difference arises because $t_{\rm m}$ is an integrated quantity that depends on the full orbital evolution of the binary.
For all binaries shown in Fig.~\ref{merger_time}, with the exception of the two most eccentric VG+25 systems discussed below, the maximum (minimum) merger times are obtained at $f \approx \pi/2$ ($f \approx 3\pi/2$), in agreement with Ref.~\cite{2025PhRvD.112b4012F}. These two points correspond approximately to the beginning and the end of the pericenter region, namely the portion of the orbit where GW emission is strongest. Initializing a binary just before entering this region implies that it experiences an intense burst of GW emission almost immediately. Conversely, initializing the binary just after it, places the BHs at the point along the orbit that is farthest from the next episode of strong GW emission.%

Crucially, this interpretation requires that the binary does not undergo significant secular evolution before reaching the next pericenter. In particular, for an initial true anomaly $f_0$, the time to the next pericenter, $\Delta t_{\rm peri}(f_0)$, should remain shorter than the ZLK timescale,
$\Delta t_{\rm peri}(f_0) < \tau_{\rm ZLK}$
Otherwise, the binary may undergo appreciable ZLK evolution before the next episode of strong GW emission, and the assumption that its eccentricity evolves solely due to GW emission during this interval is no longer valid. We therefore impose this condition when identifying the extrema of the merger time.

This consideration is particularly relevant for the two most eccentric binaries in the VG+25 sample. For the most eccentric binary, the maximum (minimum) value of $t_{\rm m}$ is obtained for an initial true anomaly $f\simeq2.21$ ($f\simeq4.45$), whereas for the second most eccentric system the maximum (minimum) occurs at $f\simeq 1.685$ ($f\simeq4.79$).

The difference between these systems and the other binaries arises for two reasons. First, after mapping the initial conditions to ($\bar e_{\rm in},\bar p_{\rm in}$), we find $\bar e_{\rm in}>1$ for certain values of $f$, corresponding to hyperbolic orbits. Consequently, not all initial values of the true anomaly lead to a finite merger time. We discuss this effect in Appendix~\ref{hyperbolic}.

Second, for these highly eccentric systems, some of the initial phases that would otherwise yield the longest merger times correspond to $\Delta t_{\rm peri}>\tau_{\rm ZLK}$. Including such phases would therefore violate the assumption that the binary can be treated as evolving in isolation until the next pericenter passage. We consequently exclude these phases when determining the extrema reported above. If this timescale constraint is neglected, the longest merger times for the two most eccentric binaries would instead be obtained at $f\simeq2.19$ for the most eccentric system and $f\simeq1.27$ for the second most eccentric system. 
Finally, the merger times estimated using the definition of Ref.~\cite{2018ApJ...863...68L} (see Eq.~\ref{tLL}) generally overestimate those obtained using the Peters and NA equations. The two most eccentric binaries in the VG+25 sample are notable exceptions: for these systems, the estimate of Ref.~\cite{2018ApJ...863...68L} becomes comparable to, or smaller than, the merger times obtained with the Peters and NA prescriptions. %

\section{Discussion and conclusions}\label{discussion and conclusions}

The merger time plays a crucial role in determining the relative contribution of different BH formation channels, as it directly affects the delay-time distribution of compact binaries. In this work, we investigate how the treatment of binary dynamics affects the predicted merger time for eccentric binaries. We compare merger times obtained from orbit-averaged and NA equations, and from the prescription of Ref.~\cite{2018ApJ...863...68L}, for two populations of highly eccentric binaries in hierarchical triple systems. In these systems, the ZLK mechanism can drive the eccentricity to extreme values, bringing the binary into a regime where the ratio between the radiation-reaction and orbital timescales becomes smaller than one and the usual adiabatic approximation may break down. For the NA evolution, we additionally study the dependence of the merger time on the initial orbital phase by varying the initial true anomaly.

We find that the orbit-averaged description remains remarkably accurate over a broad range of eccentricities, even for systems satisfying $\tau_{\rm rr}\ll\tau_{\rm orb}$. Significant discrepancies with the NA evolution arise only for the most extreme configurations, characterized by simultaneously large eccentricities and small semi-latus recta. However, comparison with the prescription of Ref.~\cite{2018ApJ...863...68L} further shows that the choice of merger-time approximation can introduce systematic differences, with this prescription generally overestimating the merger time relative to both the orbit-averaged and NA calculations.

Our results show that the agreement between the orbit-averaged and NA descriptions depends strongly on the initial conditions of the binary. For the VG+25 population, the two approaches give consistent merger times for moderately eccentric systems, whereas significant discrepancies arise for the most eccentric binaries, for which the choice of initial true anomaly plays a substantial role. In contrast, the L\&L18 population, despite reaching even larger eccentricities, generally shows agreement between the two approaches, because its binaries start at larger separations and therefore undergo weaker GW dissipation.

The origin of these discrepancies lies in the interplay between the strongly phase-dependent nature of GW emission and the assumptions underlying orbit averaging.

Gravitational-wave emission from eccentric binaries is highly asymmetric along the orbit. This can be appreciated by considering the instantaneous energy and orbital angular-momentum fluxes, which exhibit a strong dependence on the binary separation \cite{1995PhRvD..52.6882I}. At leading quadrupole order, the fluxes scale approximately as
\begin{align}
\frac{dE}{dt}\propto r^{-5},\,\,\,\,\,
\frac{dL}{dt}\propto r^{-4},
\end{align}
where
\begin{align}
r=\frac{p}{1+e\cos f}.
\end{align}
Thus, the GW emission is strongly enhanced during the short passage through pericenter, where the binary separation is smallest, while the contribution from the long apocenter passage is much smaller.
Orbit averaging does not neglect this asymmetric emission. It integrates the instantaneous dissipation over one complete orbital period, weighting each orbital phase by the time spent at the corresponding $r$, and replaces the rapidly varying emission with a single orbit-averaged dissipation rate. This procedure is justified only if the orbital elements change negligibly during one orbit. That is, if the binary evolves adiabatically and the orbit can be treated as approximately fixed over the averaging timescale.

When this assumption holds, the influence of the initial true anomaly on the merger time is minimal. Although GW emission formally breaks time-reversal symmetry \cite{poisson}, the energy dissipated during a single orbit is small enough that two binaries initialized at different orbital phases remain nearly identical. In practice, a binary initialized just after pericenter passage ($f\sim\pi/2$) simply needs to complete almost one additional orbit before experiencing the same strong burst of GW emission as a binary initialized just before pericenter passage ($f\sim 3\pi/2$); their inspirals therefore differ only by a small time offset and can be regarded as time-shifted copies of each other. Because the binaries complete many orbital cycles before the eccentricity and semi-latus rectum change appreciably, the evolution is effectively adiabatic and orbit-averaged equations accurately describe the inspiral.

When the eccentricity becomes sufficiently large and the semi-latus rectum sufficiently small, a large fraction of the orbital energy is radiated during a single pericenter passage. In this regime, the orbital elements evolve significantly within one orbit, violating the fundamental assumption of orbit averaging. Two binaries initialized just before or after pericenter no longer represent simple time-shifted versions of the same inspiral: the one initialized just before pericenter undergoes its first intense burst of GW emission almost immediately, while the other must travel for approximately $3/4$ of an orbital period to reach the same position. By then, contrary to what is assumed in orbit averaging, the first binary has already evolved onto a different orbit, with different values of the semi-latus rectum and eccentricity. The two systems therefore follow genuinely different dissipative trajectories, resulting in significantly different merger times. This behavior explains the broad range of merger times obtained with the non-orbit-averaged equations shown in Fig.~\ref{merger_time}.

In this regime, orbit averaging ceases to provide a meaningful description of the dynamics, and large discrepancies with the non-orbit-averaged predictions are observed.%

Nevertheless, for most systems, these differences do not qualitatively affect the astrophysical interpretation of the merger time. The situation changes only for the most extreme binaries: in some configurations, accounting for the non-orbit-averaged dynamics changes the predicted merger time from a timescale of years to only a few seconds.

These results have direct implications for population studies of eccentric compact binaries. Since merger times directly determine the relative contribution of different BH formation channels, one would expect that the extreme orbital conditions reached by these systems generally demand a non-orbit-averaged treatment when their merger times are used to infer astrophysical origin. However, we find that the two descriptions remain comparable for the majority of cases, and the robustness of the orbit-averaged merger times suggests that they remain a reliable and computationally efficient approximation for population-level studies. The most extreme binaries instead represent a potentially important tail where the details of the dynamical treatment can qualitatively alter the inferred delay-time distribution.

Finally, it is important to emphasize that our conclusions concern quantities that are integrated over the binary inspiral. Our results should not be interpreted as implying that orbit-averaged equations provide an equally accurate description of instantaneous dynamical quantities. In applications that require an accurate description of the binary dynamics on orbital timescales, such as waveform modeling or the evolution close to merger, the non-orbit-averaged dynamics may become important even when the resulting difference in the total merger time is small.

\acknowledgements

We thank Andrew Chael, Davide Gerosa, Evgeni Grishin, Konstantinos Kritos, Nick Loutrel, Johan Samsing, Brian Seymour, Jakob Stegmann and Saul Teukolsky for discussions. 
G.F. acknowledges the support of the European Consortium for Astroparticle Theory in the form of an Exchange Travel Grant.
G.F. is supported by the Sherman Fairchild Postdoctoral Fellowship at the California Institute of Technology.
A.V.-G. is supported by the Carlsberg Foundation, grant CF24-1046.

\bibliography{ZLKpaper}

\appendix
\section{Hyperbolic case}\label{hyperbolic}

 \begin{figure*}[h!]
	\includegraphics[width=0.87\textwidth]{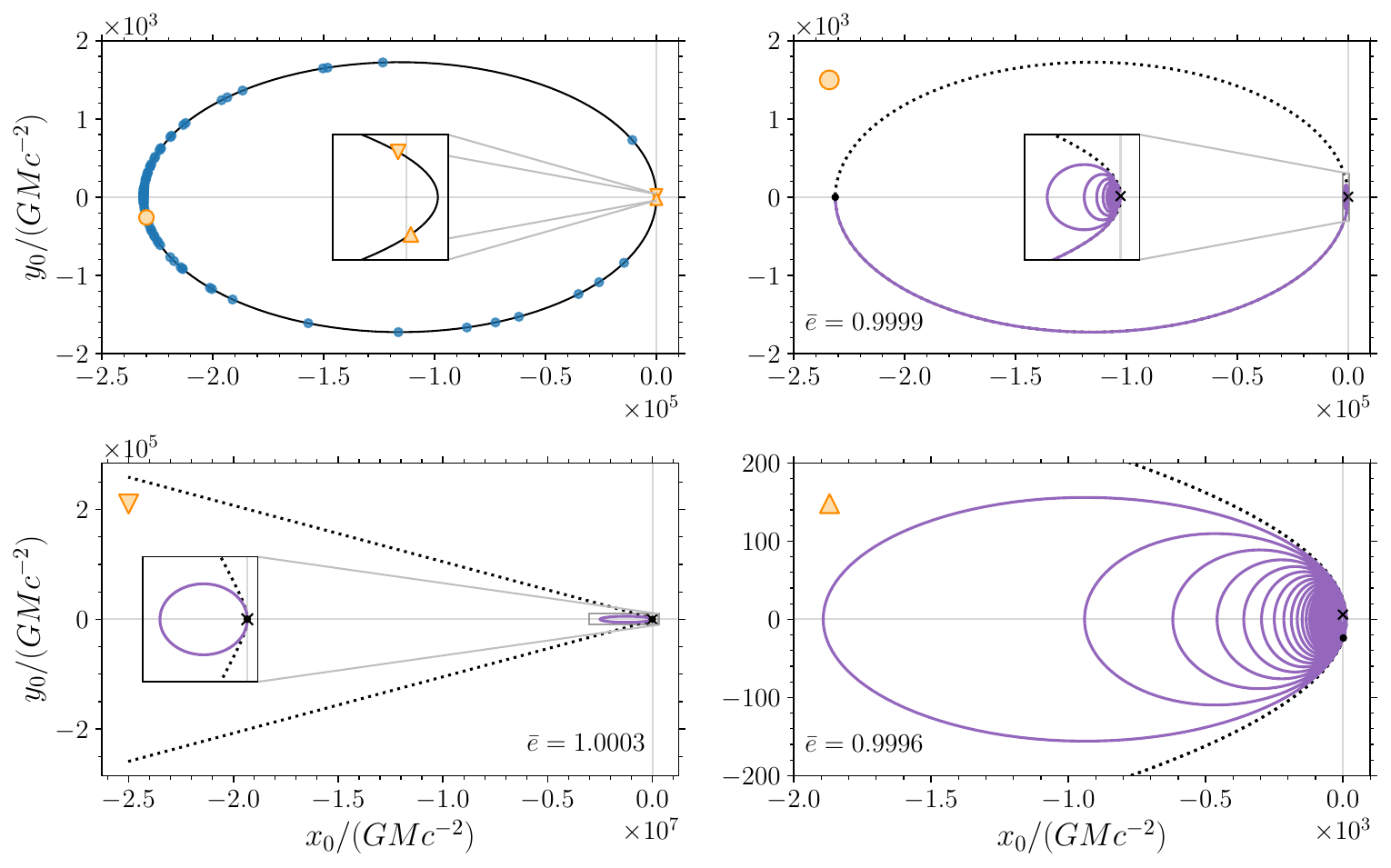} 
	\caption{ Upper-left panel: sampled values of the true anomaly $f$ (blue circles) and the allowed extrema (orange triangles) used to estimate the merger time $t_{\rm m}$ with the non-orbit-averaged (NA) equations for the second most eccentric binary in the VG+25 population considered in this work. Upper-right panel: orbital evolution obtained with the NA equations for one of the sampled values of $f$ (orange circle), whose merger time is representative of the mean evolution. Bottom-left (bottom-right) panel: evolution corresponding to the value of $f$ that yields the longest (shortest) merger time. In all evolution panels, the purple curves show the orbit-averaged evolution parametrized by $\bar{e}$, $\bar{p}$, and $\bar{t}$. Black dots mark the initial conditions, while crosses indicate the end of the evolution.}
	\label{3evolution}
\end{figure*} 
\begin{figure*}[h!]
	\includegraphics[width=0.87\textwidth]{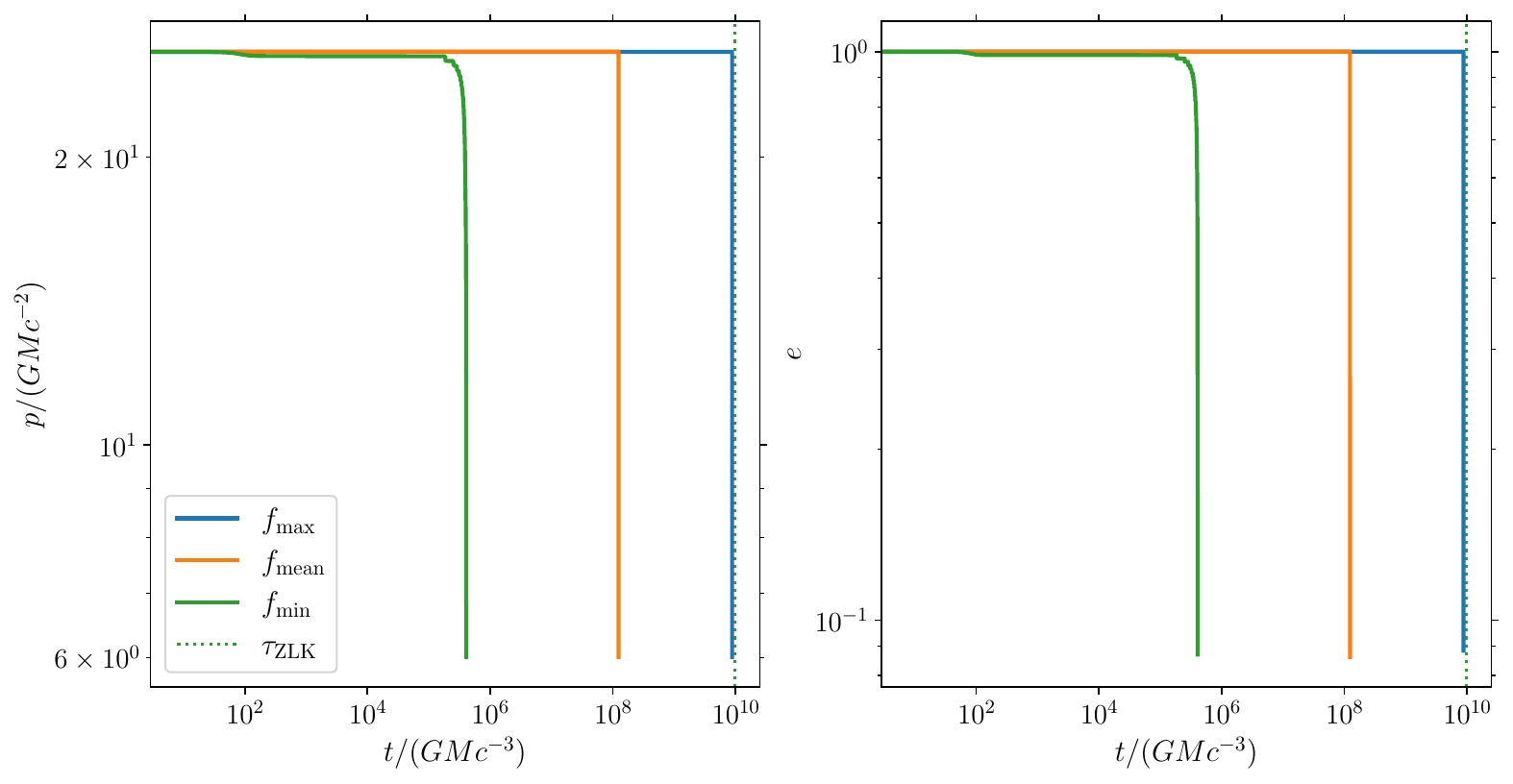} 
	\caption{ Evolution of the semi-latus rectum $p$ (left panel) and eccentricity $e$ (right panel) as functions of time, obtained by integrating the NA equations for the second most eccentric binary in the VG+25 population considered in this work. Each curve corresponds to a different choice of the initial true anomaly $f$. The green (blue) curve shows the evolution for the value of $f$ that yields the minimum (maximum) merger time, while the orange curve represents the mean evolution obtained by averaging over 100 sampled values of $f$, weighted according to their probability.  }
	\label{pande}
\end{figure*} 

We report in Figs.~\ref{3evolution} and~\ref{pande} three representative orbital evolutions of the BH binary belonging to the VG+25 population with initial astrophysical parameters ($p_0=25.8\,GM/c^2,e_0=0.9999$), corresponding to the second most eccentric system shown in the left panel of Fig.~\ref{merger_time} for three different choices of the initial true anomaly. As discussed in Sec.~\ref{conversions}, these initial conditions must first be converted into the variables required by the NA formalism. Consequently, each value of the true anomaly $f$ corresponds to a different set of initial conditions in the NA evolution.

The value of $f$ that maximizes the merger time $t_{\rm m}$ for the binary considered here corresponds to an initially unbound configuration ($\bar{e}_{max}=1.0004$). The binary therefore begins on a hyperbolic orbit, but the two BHs pass sufficiently close to one another that GW emission rapidly removes enough energy and angular momentum to produce a bound system with eccentricity $(1-e)\sim10^{-5}$. After becoming bound, however, the binary finds itself just after pericenter passage. Since the evolution proceeds toward increasing true anomaly, the BHs must complete almost one full orbit before reaching the pericenter region again ($\Delta t_{\rm peri}^{\rm max}\sim 8.8\times10^{9} GM/c^3$). During this time, the orbital elements remain almost unchanged because the GW emission is negligible (Fig.~\ref{3evolution}). Once the binary reaches pericenter again, the small separation produces an intense burst of GW emission, rapidly driving the system to merger.

On the other hand, the value of $f$ that produces the shortest merger time corresponds to an initially bound, highly eccentric binary that is already approaching the pericenter region ($\Delta t_{\rm peri}^{\rm min}\sim 78 GM/c^3$). Consequently, the binary experiences an intense burst of GW emission almost immediately, causing both the separation and the eccentricity to decrease rapidly. Although the system subsequently completes many short orbital cycles before merging, the inspiral is significantly shorter than in the previous case because the first strong dissipative episode occurs almost immediately.

Finally, for the most probable initial configuration $f\simeq\pi$, the binary starts just after apocenter on a highly eccentric bound orbit. It must therefore complete approximately half an orbital period before reaching the pericenter region ($\Delta t_{\rm peri}^{\rm mean}\sim 1.2 \times 10^{8} GM/c^3$) and undergoing its first significant burst of GW emission. From that point onward, the evolution closely resembles the previous case, with the binary rapidly shrinking during successive pericenter passages until merger.

\end{document}